\documentclass[natbib]{llncs}
\usepackage{amsmath}
\usepackage{amssymb}
\usepackage{stmaryrd} 
\usepackage{pifont}
\usepackage{fancyvrb}
\usepackage{graphicx}
\usepackage{enumerate}

\title{A Web Interface for Matita}

\author{Andrea Asperti \and Wilmer Ricciotti}

\institute{Department of Computer Science, University of Bologna\\
           \email{\{asperti,ricciott\}@cs.unibo.it}}

\begin{document}
\maketitle

~\vspace{0.5cm}~

\noindent
This article describes a prototype implementation of a web interface
for the Matita proof assistant~\cite{Matita-cade}. The motivations behind our
work are similar to those
of several recent, related efforts \cite{web-interfaces,wiki-mizar,large-wikis,proviola}
(see also~\cite{geuvers}). In particular:
\begin{enumerate}
\item creation of a web collaborative working environment for interactive
theorem proving, aimed at fostering knowledge-intensive cooperation, content 
creation and management;
\item exploitation of the markup in order to enrich the document
with several kinds of annotations or active elements; annotations may have both
a presentational/hypertextual nature, aimed to improve the quality of the proof
script as a human readable document, or a more semantic nature, aimed to help the system in its processing (or re-processing) of the script;
\item platform independence with respect to operating systems, and wider accessibility also for users using devices with limited resources;
\item overcoming the installation issues typical of interactive provers, 
also in view 
of attracting a wider audience, especially in the mathematical community.
\end{enumerate} 

Point 2. above is maybe the most distinctive feature of our approach, 
and in particular the main novelty with respect to \cite{web-interfaces}.

In fact, delivering a proof assistant as a web application enables us to 
exploit the presentational capabilities of a web browser with little effort.
Purely presentational markup does not require any special treatment on
the part of the prover and is natively supported by the web browser.
However, having an easy access to HTML-like markup allows much more flexibility.
Not only can we decorate comments by means of textual formatting or pictures;
executable parts of scripts reference concepts defined elsewhere, either in the
same script or in the library, using possibly overloaded identifiers or
notations: it is natural to enrich those identifiers with hyperlinks to the 
associated notions. This association is actually computed by the 
system every time the script is parsed, hence it is the system's job to 
enrich the script accordingly.
Since the previous computation can be expensive, it is natural to have
the system use such hyperlinks to speed up the execution of the script.
Moreover, when the source text is particularly ambiguous, hyperlinks provide 
essential semantic information to avoid asking the user for explicit
disambiguation every time the script is executed. 

Hyperlinks are an example of a textual annotation having both a presentational
and a semantic value. The text enriched with hyperlinks not only provides a
more dynamic and flexible format to access the repository, but it is also
a more explicit and hence more robust representation of the information.

A further use of markup is to attach to the script information
that is valuable to the system, but is not thought to be normally read by the
user. This is technically a kind of presentational markup, used to hide parts of
the script rather than for decorating text.

Our current implementation supports three categories of markup:
\begin{itemize}
\item standard HTML markup, used to add formatting to text;
formatted text is currently assumed to occur in Matita comments;
\item hyperlinks to Matita definitions, typically produced by the system and
reused on a new parsing of the script to avoid a second
disambiguation of the input (at the time of the submission, traversing 
hyperlinks is not yet supported, but implementing it does not look 
problematic);
\item markup wrapping traces of execution of automation steps in the script,
produced by the system on a first execution and granting a notable speed-up
on future executions; the trace is normally transparent to the user, but visible
on demand. 
\end{itemize}

\section*{Structure of the system}

\paragraph{Matita core}
The server runs a minimally reworked version of the Matita engine, equivalent to
its stand-alone counterpart, but for the following features:
\begin{itemize}
\item the status of Matita includes the user id of its owner, as needed by an
inherently multi-user web application: this allows the system to run at the same
time several user-specific versions of the library;
\item the lexical analyzer and the parser take into account the script markup;
\item the disambiguation engine and the automation tactic produce and return
information suitable for enriching the script.
\end{itemize}

For what concerns the lexical analyzer, producing specific tokens for the markup
would require major modifications to the parser, which in Matita is a complex
component extensible at runtime with user provided notations. In an effort to
keep the parser as untouched as possible, the token stream returned by our
lexical analyzer ignores the markup; however, hyperlinks that can
be used for immediate disambiguation are stored in an additional table that is
later accessible to the parser, which is then able to build a disambiguated 
abstract syntax tree (AST)
for it. In order for this technique to work, we assume that disambiguation
markup is only located around ``leaves'' of the AST (and in particular,
identifiers or symbols); at the moment, this assumption does not seem to be
restrictive. 

Markup for automation traces, which is used only to hide additional arguments to
the automation tactic, is completely handled by the user interface and can thus
be safely ignored by the lexical analyzer and the parser.

\paragraph{Matita web daemon}
The Matita web daemon is a specialized HTTP server, developed using the
\verb+Netplex+ module of the Ocamlnet library\footnote{\url{http://projects.camlcity.org/projects/ocamlnet.html}}, providing remote access
to the Matita system. It exports several services:
\begin{itemize}
\item storage of user accounts and authentication;
\item storage of user libraries;
\item synchronization of user libraries with the shared library via \verb+svn+;
\item remote execution of scripts.
\end{itemize}
Such services are invoked through a CGI interface and return XML
documents encoding their output.

Remote execution of scripts allows a user authoring a script on a web browser to
send it to the server for processing. The typical interactions with a script are
allowed, in the style of Proof-General~\cite{proofgeneral} and similarly
to~\cite{web-interfaces}: executing one step (tactic or directive) or the whole
script, as well as undoing one step or the whole script (execution of a script
until a given point is reached is performed by the client by multiple calls to
single-step execution).

Parsing of the script is performed on the server, as client-side parsing of
the extensible syntax used by Matita is essentially unfeasible. To execute (part
of) a script, the server needs thus to receive all of the remaining text to be
parsed, because the end of the next statement is not predictable without a full
parsing. The Matita daemon will answer such a request by returning to the
client
\begin{itemize}
\item the length of the portion of the original script that has been
successfully executed;
\item a (possibly empty) list of parsed statements, which have been enriched
with mechanically generated markup including disambiguation hints and automation
traces (the length of this updated text does \emph{not} match the previous value
in general);
\item an HTML representation of the proof state of the system after the
execution of the last statement (if the execution stopped in the middle of a
proof);
\item a representation of the error that prevented a further execution of the
script (if the execution stopped because of an error).
\end{itemize}

\paragraph{Collaborative formalization}
The daemon provides a preliminary support for collaborative formalization,
currently coming in the form of a centralized library maintained by means of
$\mathsf{svn}$. Other authors (see~\cite{large-wikis}) have advocated the use of
distributed versioning systems (e.g.~$\mathsf{Git}$). Our choice is mainly
related to the reuse of the original Matita repository and to the fact that
$\mathsf{svn}$ already supports the kind of distributed activity we have in
mind. The effective usability and scalability of this approach will be tested in
the future.

\paragraph{The client} 
The Matita web client was initially written in plain Javascript and is currently
being ported to the jQuery\footnote{\url{http://jquery.com}} framework. The client implements a user interface
that is essentially similar to the one of ProofGeneral~\cite{proofgeneral},
CtCoq and CoqIDE~\cite{ctcoq3}, or stand-alone
Matita~\cite{matita-jar-uitp}, but in the form of a web page. This includes displaying the script
(disabling editing for the already executed part), buttons for script
navigation, boxes for proof state (including multiple open goals) and
disambiguation, instant conversion of \TeX{}-like escapes to Unicode symbols,
and essential interface for accessing the remote file system.

The implementation issues are similar to those described in \cite{Wenzel11}.
The web interface does not need to understand much of Matita:
information like being in an unfinished proof or in disambiguation mode can be
easily inferred from the data structures returned from the server. On the other
hand, some code is necessary to convert Matita markup to HTML markup and
vice-versa.

\paragraph{Availability}
The Matita web interface is accessible from the website
\linebreak \url{http://pandemia.helm.cs.unibo.it/login.html}. Accounts for accessing
the interface are provided by the authors on request.


\bibliographystyle{plain}
\bibliography{../../BIBTEX/helm}

\begin{thebibliography}{10}

\bibitem{large-wikis}
Jesse Alama, Kasper Brink, Lionel Mamane, and Josef Urban.
\newblock Large formal wikis: Issues and solutions.
\newblock In {\em Proceedings of Intelligent Computer Mathematics (CICM 2011),
  Bertinoro, Italy}, volume 6824 of {\em Lecture Notes in Computer Science},
  pages 133--148. Springer, 2011.

\bibitem{Matita-cade}
Andrea Asperti, Wilmer Ricciotti, Claudio~Sacerdoti Coen, and Enrico Tassi.
\newblock The {M}atita interactive theorem prover.
\newblock In {\em Proceedings of the 23rd International Conference on Automated
  Deduction (CADE-2011), Wroclaw, Poland}, volume 6803 of {\em LNCS}, 2011.

\bibitem{matita-jar-uitp}
Andrea Asperti, Claudio {Sacerdoti Coen}, Enrico Tassi, and Stefano Zacchiroli.
\newblock User interaction with the {M}atita proof assistant.
\newblock {\em Journal of Automated Reasoning}, 39(2):109--139, 2007.

\bibitem{proofgeneral}
David Aspinall.
\newblock {P}roof {G}eneral: A generic tool for proof development.
\newblock In {\em Tools and Algorithms for the Construction and Analysis of
  Systems, TACAS 2000}, volume 1785 of {\em Lecture Notes in Computer Science}.
  Springer-Verlag, January 2000.

\bibitem{ctcoq3}
Yves Bertot and Laurent Th\'ery.
\newblock A generic approach to building user interfaces for theorem provers.
\newblock {\em Journal of Symbolic Computation}, 25:161--194, 1998.

\bibitem{geuvers}
Herman Geuvers.
\newblock Proof {A}ssistants: history, ideas and future.
\newblock {\em Sadhana}, 34(1):3--25, 2009.

\bibitem{web-interfaces}
Cezary Kaliszyk.
\newblock Web interfaces for proof assistants.
\newblock {\em Electr. Notes Theor. Comput. Sci.}, 174(2):49--61, 2007.

\bibitem{proviola}
C.~Tankink, H.~Geuvers, J.~McKinna, and F.~Wiedijk.
\newblock Proviola: A tool for proof re-animation.
\newblock In {\em Proceedings of AISC 2010, Heidelberg}, volume 6167 of {\em
  Lecture Notes in Computer Science}, pages 440--454. Springer, 2010.

\bibitem{wiki-mizar}
Josef Urban, Jesse Alama, Piotr Rudnicki, and Herman Geuvers.
\newblock A wiki for mizar: Motivation, considerations, and initial prototype.
\newblock {\em CoRR}, abs/1005.4552, 2010.

\bibitem{Wenzel11}
Makarius Wenzel.
\newblock Isabelle as document-oriented proof assistant.
\newblock In {\em Intelligent Computer Mathematics - 18th Symposium, Calculemus
  2011, and 10th International Conference, MKM 2011, Bertinoro, Italy, July
  18-23, 2011. Proceedings}, volume 6824 of {\em LNCS}, pages 244--259, 2011.

\end{thebibliography}

%
%
\end{document}